# Gauging Growth: AGI's Mathematical Metrics for Economic Progress

Davit Gondauri[1], Maia Noniashvili[1], Mikheil Batiashvili[1] & Nino Enukidze[1]

[1] Business and Technology University, Tbilisi, Georgia

Correspondence: Davit Gondauri, Business and Technology University, Tbilisi, Georgia. E-mail: dgondauri@gmail.com

**Abstract**

Today, the economy is greatly influenced by Artificial General Intelligence (AGI). The purpose of this paper is to determine the impact of the quantitative relations of AGI on the country's economic parameters. The authors use the analysis of historical data in the research, develop a new mathematical algorithm that refers to the level of AGI development, and conduct the regression analysis. The economic effect of AGI is deduced if it affects the growth of real GDP. As a result of the analysis, it is revealed that there is a positive Pearson correlation between the growth of AGI and real GDP, that is, to increase GDP by 1%, an average increase of 12.5% of AGI is required.

**Keywords:** Artificial General Intelligence (AGI), assessment of AGI impact, regression coefficient

## 1. Introduction

AGI is of great importance in the evolution of technology. It can also give us a clear picture of how the economy works. Today, when AGI is becoming an integral part of the socioeconomic sphere, traditional approaches to econometrics are facing major challenges. The study attempts to solve this problem by proposing a new metric that assesses the level of AGI development and its impact on economic growth.

Background of the Research Problem: Traditional econometrics, exemplified by Gross Domestic Product (GDP), have long been the cornerstone for assessing a nation's economic health. However, the rapid advancements in AGI technologies bring forth a host of complexities that these conventional metrics struggle to encapsulate. AGI's influence extends beyond mere productivity gains, infiltrating sectors such as labor, innovation, and market dynamics. Consequently, the inadequacy of existing metrics to capture the multifaceted impact of AGI becomes a significant obstacle in understanding the true nature of economic growth.

As AGI contributes to unprecedented shifts in the workforce, alters production methodologies, and stimulates innovation at an unparalleled pace, it is imperative to reassess our tools for measuring economic progress. The research problem at hand is thus twofold: firstly, to identify and comprehend the limitations of traditional metrics in the face of AGI, and secondly, to introduce a set of mathematical metrics that can better gauge the intricate dimensions of economic growth in this transformative era.

The main purpose of the study is to examine AGI's Mathematical Metrics for Economic Progress. Based on the purpose of the research, we set the following tasks:

Develop Mathematical Metrics Framework:

- Devise a comprehensive framework of mathematical metrics tailored to the intricacies of economic progress in the AGI era.

- Consider and integrate variables such as technological advancements, shifts in labour markets, and the overarching influence of AGI on diverse economic sectors.

Quantify AGI's Impact on Economic Indicators:

- Utilize the developed mathematical metrics to quantify and analyse the influence of AGI on key economic indicators.

- Explore how AGI contributes to or transforms traditional economic indicators, including but not limited to productivity, employment patterns, and overall economic growth.

Assess Predictive Capabilities:

- Evaluate the predictive capabilities of the proposed mathematical metrics in forecasting economic trends amidst the integration of AGI.
- Examine the adaptability of these metrics to dynamic changes influenced by AGI technologies and their effectiveness in predicting shifts in economic patterns.

Significance of the Study:

This study holds substantial significance in providing a nuanced understanding of economic growth in the context of AGI integration. The outcomes are poised to inform policy decisions, guide economic planning, facilitate informed business strategies, and contribute valuable insights to the academic discourse surrounding the intersection of AGI and economics. By addressing the limitations of traditional metrics and introducing a more adaptive framework, this research aims to equip decision-makers with the tools needed to navigate the complexities of economic progress in the era of AGI.

## 2. Literature Review

Artificial General Intelligence (AGI) is a pivotal and contentious concept within the domain of computational research, denoting an artificial intelligence system possessing capabilities commensurate with or surpassing those of a human across a diverse array of tasks. As Machine Learning (ML) models undergo fast advancements, the contemplation of AGI has transitioned from the realms of philosophical discourse to immediate practically significant arena. Notably, some authorities posit the emergence of AGI "sparks" in modern large language models (LLMs), with assertions that the latest iterations of such models may already nurture elements reminiscent of AGI (Bubeck et al., 2023). Forecasts within the scientific community start from predictions that artificial intelligence will greatly surpass human performance within a decade (Bengio et al., 2023) to claims positing the current status of some LLMs as AGIs (Agüera y Arcas, & Norvig, 2023). In response to these evolving assumptions, scholars advocate for the creation of a structured framework for the classification of capabilities and behavioral attributes inherent to Artificial General Intelligence (AGI) models and their antecedents. (Morris M.R., et.al "Levels of AGI: Operationalizing Progress on the Path to AGI, 2023").

The integration of technologies designed to improve the production, distribution, and consumption of goods and services, as well as to stimulate the transformation of markets, plays a significant role in shaping capitalist societies. Furthermore, these technologies contribute to the refinement of the spatial division of labor. This observation is underscored in Leon's work on "AI and the capitalist space economy" (2021).

In the examination of the paradigm of artificial intelligence (AI) and its economic implications, economists Acemoglu and Restrepo (2020) contend that the replacement of human workers by AI represents just one among many scenarios. Leon (2021) further elaborates on this perspective in the context of the capitalist space economy. A big factor in forecasting technology's impact on jobs, consumers, and geopolitics is recognizing that earth-shattering progress enabled by each revolutionary technology occurs within a complex ecosystem. Moreover, its ultimate impact is shaped by the free market interplay of ecosystem participants creating an avalanche of innovations and wealth. That process obviously depends on competition among an initial set of industry participants, but it is further enhanced by the rise of new entrants and the impact of direct substitutes. At the same time, the industry benefits as its suppliers offer better inputs, and its customers develop more advanced needs. (Trends eMagazine, October 2023).

In their paper titled "Artificial Intelligence and Economic Growth" (2017), P. Aghion, B. Jones, and C. Jones searched through the implications of artificial intelligence (AI) on economic growth. The authors scrupulously examined four principal facets of the impact of Artificial General Intelligence (AGI): Automation of Production, the Idea Production Function, Singularities, and the intersection of A.I. with Firms and Economic Growth.

The investigation set by introducing A.I. into the production function of goods and services, with a particular focus on reconciling the developing landscape of automation with the observed stability in both capital share and per capita GDP growth throughout the past century. Authors' model incorporated Baumol's "cost disease" insight into Zeira's automation model, resulting in a spectrum of potential outcomes. The authors derived conditions under which balanced growth could occur, maintaining a constant capital share that remains below 100%, even in scenarios of complete automation.

Subsequently, the paper shifted its attention to the speculative hypothesis of the effects of integrating A.I. into the production technology governing the creation of new ideas. The introduction of A.I. in this context, as authors say, holds the potential to exert a variable impact on economic growth, either in a transient or enduring manner contingent upon the specific way it is introduced.

Korinek and Stiglitz (2021) have explored different variables and metrics within the context of a competitive economy. They propose that technological progress will influence the isoquants which represent the combinations of inputs—specifically, capital (K) and labor (L)—required to produce a given output. The authors argue that progress leads to an inward shift of isoquants, indicating a less need for inputs to achieve a specified output level.

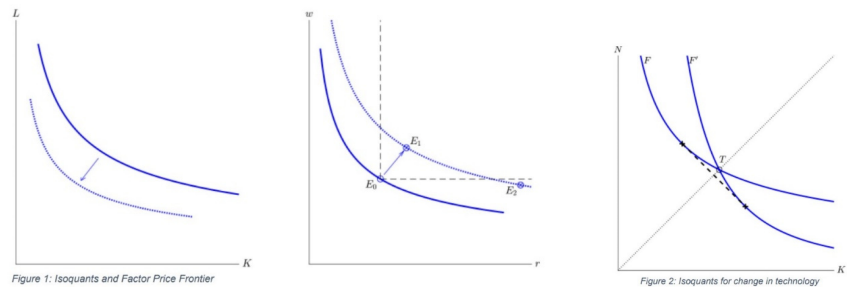

Figure 1: Isoquants and Factor Price Frontier

Figure 2: Isoquants for change in technology

Within the framework of a competitive economy, Korinek and Stiglitz introduce a straightforward criterion for measuring the impact of innovation on wages, distinguishing between scenarios where wages increase (as in $EE1$) or decrease (as in $EE2$). This criterion depends on whether, at the initial wage level ($ww0$), there is an increase or decrease in the demand for labor resulting from the technological progress. The authors write that the distributive effects of innovations create quasi-rents, with winners (e.g., capitalists or skilled workers) benefiting from progress without automatically contributing to the innovation, while losers experience comparable losses. While the article acknowledges the implications of resource-saving technological progress in the context of Artificial Intelligence (AI), this specific topic is not explored further. In this regard it is important to note authors' assertion that recent technologies may have generated increases in societal welfare that traditional metrics, such as Gross Domestic Product (GDP), fail to fully capture. They point out the instances where services are exchanged for "eyeballs," signifying that users are exposed to advertisements rather than paying for services. This perspective aligns with a broader view on the limitations of GDP in capturing the multifaceted impacts of modern technological advancements, as shown by the work of Brynjolfsson (2020).

Building upon the insights gleaned from the literature review we conducted, our research aims to delve deeper into the subject matter, employing mathematical tools to elucidate the intricacies of Artificial General Intelligence (AGI). Khalili's article (2021) assumes special significance in this pursuit, offering a distinctive perspective on the role of mathematics in comprehending AGI. Khalili draws parallels between the application of mathematical tools in AI and their utility in diverse scientific fields. In the context of globalization and economic development, Korinek and Stiglitz (2021) conclude by emphasizing the potential of economic analysis, based on models suitable for the AGI era, in shaping global and national policies. The goal is to mitigate adverse effects and ensure that AI-driven innovation contributes to increased standards of living, especially in developing countries.

## 3. Methodology

The research methodology includes developing a metrics, quantifying the impact of AGI on economic performance, and evaluating the predictive capabilities of the proposed metrics.

Based on the literature review, we identify the relevant variables and concepts in the relationship between economic growth and AGI. Then we collected various data sets from official sources on the Internet, which include the following indicators: Labor Force Productivity, Total factor productivity index, Capital Stock, Real Gross Domestic Product, and output elasticity of capital (Cobb-Douglas).

In the next step, we used a developed mathematical algorithm to calculate the level of development of AGI and its impact on key economic indicators. After developing the AGI development level model, regression analysis was conducted to assess the relationships between economic indicators and AGI-based variables.

Mathematical Model: Economic Effects of AGI on Economic Growth

Variables:

- $Y_n$ - Real GDP at time n.
- $K_n$ - Capital stock at time n.
- $L_n$ - Labor force at time n.
- $A_n$ - Total factor productivity (TFP) at time n.
- $AGI_n$ - The level of AGI technology at time n.

- $E_n$ - Total labor force productivity at time n.

Assumptions and Explanations:

1. We used the Solow Growth Model as the basis of the research, which is based on the aspects of the economic growth theory. It assumes that economic growth is determined by labour force growth, technological progress, and capital accumulation.

2. Total factor productivity (TFP) compares total outputs relative to the total inputs used in the production of the output. As both output and inputs are expressed in terms of volume indices, the indicator measures TFP growth. AGI has the potential to increase TFP through innovation, automation, and optimization, making production more efficient, and improving existing technologies.

3. AGI Impact on TFP: The level of AGI technology, denoted as $AGI_n$, represents the extent to which AGI has been integrated into the economy. As $AGI_n$ increases, it positively affects A by boosting TFP. The higher the $AGI_n$, the greater the potential for economic efficiency and innovation.

4. Labor Force Productivity: Et represents the overall productivity of the labour force, which includes both human and AGI labour. As AGI becomes more sophisticated and capable, it can complement human labour and increase $E_n$.

Equations:

Output (GDP):

$$Y_n = A_n \times K_n^\alpha \times (L_n \times AGI_n)^{(1-\alpha)} \tag{1}$$

$Y_n$ is the real GDP.

$K_n$ is the capital stock.

$L_n$ is the labor force.

$AGI_n$ is the AGI technology level.

$A_n$ is the total factor productivity.

α is the output elasticity of capital.

AGI technology level ($AGI_n$) formula is:

$$AGI_n = \frac{Y_n}{(A_n \times K_n^\alpha \times L_n^{(1-\alpha)})^{\frac{1}{1-\alpha}}} \tag{2}$$

Total Factor Productivity (TFP):

$$A_t = \frac{K_n^\alpha \times L_t \times AGI_n^{1-\alpha}}{Y_n} \tag{3}$$

- TFP, At, is calculated as the residual after accounting for capital and labour inputs. It represents the efficiency of converting these inputs into output.

Labor Force Productivity with AGI:

$$E_n = L_n \times AGI_n + (1 - AGI_n) \times L_n \tag{4}$$

- Labor force productivity, En, combines human labour ($L_n$) and AGI labor ($AGI_n$). It reflects the overall productivity of the labor force, considering AGI's contribution.

The output elasticity of capital (α) is the exponent in front of the capital input term. In the Cobb-Douglas production function, α represents the proportion of total output attributed to capital. It is also interpreted as the percentage change in output resulting from a 1% change in the amount of capital input, holding other factors constant. So, the formula for the output elasticity of capital (α) in the Cobb-Douglas function is:

$$\alpha = \frac{\text{Change in Output (\%)}}{\text{Change in Capital Input (\%)}} \tag{5}$$

After the obtained results, we derived the logarithms of $AGI_n$ and Real Gross Domestic Product (with Nepper base). And then we have already performed a regression analysis of the received logarithms with the following formula of the regression coefficient:

$$b_1 = \frac{\sum_{i=1}^n (x_i - \bar{x})(y_i - \bar{y})}{\sum_{i=1}^n (x_i - \bar{x})^2} \tag{6}$$

After the regression coefficient, we also calculated the following coefficients:

Pearson's correlation coefficient:

$$r = \frac{\sum(x-\bar{x})(y-\bar{y})}{\sqrt{\sum(x-\bar{x})^2 \sum(y-\bar{y})^2}} \quad (7)$$

$$\text{Coefficient of determination - } (7)^2 \quad (8)$$

Also, we used the p-value, which helps us evaluate the evidence against the null hypothesis in a statistical hypothesis test:

$$P = 2 \times P(T > |t|) \quad (9)$$

Where T is the t-distribution, and t is the t-statistic.

This comprehensive methodology aims to rigorously address the research problem, fulfill research objectives, and provide valuable insights into the complex dimensions of economic growth in the AGI era. The combination of quantitative and qualitative approaches provides a robust and multifaceted analysis that contributes to a broader understanding of the impact of AGI on economic progress.

**4. Result and Discussion**

The results of our research come from several directions. First - using the Solow model of economic growth, we examined the formula of AGI level of technology ($AGI_n$) on the example of the USA – $AGI_n = \frac{Y_n}{(A_n \times K_n^\alpha \times L_n^{(1-\alpha)})^{\frac{1}{1-\alpha}}}$.

Second, we conducted a regression analysis of AGI level of technology ($AGI_n$) and real GDP, extracting the following details. The results of linear regression analysis provide valuable insights into the relationship between variables and the overall fit of the model.

As a result of the research, it was established that:

1. The regression coefficient is 12.5%.

2. The coefficient of determination is 73%.

3. Correlation coefficient 85.4%.

4. The P-value is 0.0004123.

The regression coefficient:

The regression coefficient of 12.5% between two variables in a regression model represents the percentage change in the dependent variable for a one-unit increase in the independent variable. Let's break down the interpretation in more detail:

Definition of regression coefficient:
- The regression coefficient associated with the independent variable in a regression equation.
- It indicates the change in the dependent variable for a one-unit change in the independent variable.

Percentage Change:

A regression coefficient of 12.5% means that for every one-unit increase in the independent variable, the dependent variable is expected to change by 12.5%.

This implies a proportional relationship between the two variables.

Direction of change:

• If the regression coefficient is positive (e.g., +12.5%), it indicates a positive correlation between the two variables. If one variable increases, the other variable will increase by the specified percentage.

• If the regression coefficient is negative (e.g. -12.5%), this indicates a negative correlation. If one variable increases, the other variable of the algorithm will decrease by the specified percentage.

Coefficient of Determination:

A coefficient of determination $R^2$ with a value of 73% indicates that approximately 73% of the sensitivity of the dependent variable in the study is explained by the independent variable. A high $R^2$ value obtained implies that the presented model provides a good fit to the data.

Pearson Correlation coefficient:

A strong positive 85.4% correlation indicates a high degree of association between the independent and dependent variables.

P-value:

The low p-value of 0.0004123 obtained in the study is below conventional significance levels (e.g., 0.05). The result clearly indicates that the relationship between the variables is statistically significant.

## 5. Conclusion

The mathematical model presented by the authors is based on the Solow Growth Model, which provides a structuring framework for understanding the effects of AGI on economic growth. The algorithm exploits the dynamic interplay between AGI technology, labor, capital, and total productivity. Output elasticity of capital and regression analysis helped the analysis process.

The research findings will pave the way for informed future decision-making and will strengthen the academic discourse on the transformative impact of AGI on economic progress. Moreover, the inclusion of additional variables in future research and refinement of the model will enhance the predictive power of the algorithm.

**Acknowledgments**

We greatly appreciate the valuable contributions of our community advisory committee members. We would also like to thank Business and Technology University and every team member who took the time to contribute to this study.

**Authors contributions**

All authors contributed equally to the study, read, and approved the final manuscript.

**Competing interests**

The authors declare that they have no known competing financial interests or personal relationships that could have appeared to influence the work reported in this paper.

**Informed consent**

Obtained.

**Ethics approval**

The Publication Ethics Committee of the Canadian Center of Science and Education.

The journal and publisher adhere to the Core Practices established by the Committee on Publication Ethics (COPE).

**Provenance and peer review**

Not commissioned; externally double-blind peer reviewed.

**Data availability statement**

The data that support the findings of this study are available on request from the corresponding author. The data is not publicly available due to privacy or ethical restrictions.

**Data sharing statement**

No additional data is available.

**Open access**

This is an open-access article distributed under the terms and conditions of the Creative Commons Attribution license (http://creativecommons.org/licenses/by/4.0/).

**Copyrights**

Copyright for this article is retained by the author(s), with first publication rights granted to the journal.